\documentclass[reprint,superscriptaddress,amsmath,amssymb,twocolumn,prl]{revtex4-2}

\usepackage{amsmath}
\usepackage{mathtools}
\usepackage{amssymb}
\usepackage{graphicx}
\usepackage{bm}
\usepackage[colorlinks,linkcolor=blue,citecolor=blue,urlcolor=blue,breaklinks=true]{hyperref}
\usepackage{booktabs}
\usepackage{ragged2e}
\usepackage{orcidlink}
\usepackage[T1]{fontenc}
\usepackage{mathrsfs}

\newcommand{\controlclass}{\mathscr{C}}
\newcommand{\controlclassJ}{\mathscr{J}}
\newcommand{\figpanel}[2]{\hyperref[#1]{\ref*{#1}(#2)}}
\newcommand{\dd}{\mathrm d}

\begin{document}
\date{\today}

\title{Finite-Time Optomechanical Cooling by Multi-Exceptional-Point Braiding}

\author{Borhan Ahmadi\orcidlink{0000-0002-2787-9321}}
\email{borhan.ahmadi@ug.edu.pl}
\address{International Centre for Theory of Quantum Technologies, University of Gda\'nsk, ul.~prof. Marii Janion 4, 80-309 Gda\'nsk, Poland}

\begin{abstract}
Cooling a mechanical mode in finite time is a routing problem: excitation must reach a lossy mode before thermal noise rebuilds the population. Exceptional points are non-Hermitian degeneracies at which two hybrid modes and their eigenvectors merge, so a loop around them can exchange connected spectral branches. The main obstacle in testing whether this topology improves cooling is causal. Independently optimized enclosing and non-enclosing protocols normally have different waveforms, so topology and waveform geometry change together. We remove this ambiguity in a three-mode optomechanical system with a strongly damped main cavity and an exactly lossless auxiliary cavity. We keep one power modulation and one detuning shape fixed and change only their relative phase. The shift preserves the matched local control resources but changes the loop from non-enclosing to a braid around both exceptional points. Certified dynamics show that the two-exceptional-point protocol lowers the final mechanical occupation by more than 55 percent. The advantage survives the complete admissible phase family, deliberate waveform changes, a finite neighborhood of nearby smooth controls, and the restoration of counter-rotating heating processes. The auxiliary cavity acts as a coherent buffer, while the main cavity remains the only optical loss channel.
\end{abstract}

\maketitle

\textit{Introduction}.--
Cooling is an irreversible task. Moving excitation out of a target mode is useful only if that excitation reaches a channel from which it can be removed before thermal noise rebuilds the target population. Reservoir engineering uses this principle by shaping couplings and spectra so that unwanted excitations preferentially reach a cold loss channel \cite{Zurek2003Decoherence,Koch2016OpenQuantumControl,Breuer2016NonMarkovian,PhysRevLett.110.113604,h1fh-83td,nktx-4ltm,PhysRevLett.132.210402,ahmadi2025harnessing,ahmadi2025reservoir}. In cavity optomechanics, anti-Stokes cooling follows this route: a mechanical excitation is converted into a cavity photon, and optical decay removes the photon \cite{Aspelmeyer2014,Marquardt2007,WilsonRae2007,Vanner2013CoolingMeasurement,PhysRevLett.104.123604,Lai2021AuxiliaryCavityRefrigeration}. For a finite protocol, the timing of this route matters as well as the coupling strength. Excitation must be transferred, possibly stored, and exposed to loss at the right stage of the evolution \cite{hrph-dbv7,PhysRevLett.134.070401,PhysRevE.111.014152}.

Dissipation also changes the spectrum that organizes these transfers. The linear dynamical matrix of an open system is generally non-Hermitian, so each hybrid mode has a complex eigenvalue that contains both oscillation and decay information. Away from a degeneracy, the hybrid modes remain distinct as the controls are varied. At an exceptional point (EP), two eigenvalues and their eigenvectors merge \cite{Heiss2012,ElGanainy2018,MiriAlu2019}. The two modes then become connected sheets of a branched spectrum. A closed loop around the EP returns the controls to their starting values but can carry one continuously followed spectral sheet into the other. If the two sheets have different mixtures of mechanical and cavity character, the exchange can move excitation-bearing character from a long-lived sector to a strongly damped one. The EP does not create dissipation, and the physical state need not follow one instantaneous eigenmode adiabatically. Its role is to change the global spectral route available to the exact finite-time dynamics.

EP structure has already been used to steer states and energy in optomechanical and photonic systems. Dynamical encircling has produced topological energy transfer between mechanical modes \cite{Xu2016TopologicalEnergyTransfer}; non-Hermitian degeneracies have been used to reshape cooling, thermal fluctuations, and phonon transport \cite{Xu2019NonreciprocalCooling,Jiang2021HeatingResistantCooling,Qin2022CompensatoryCooling,Ege2024SelectiveCooling,PhysRevLett.134.043601}; and closed-loop traversal of EP branch structure is experimentally controllable \cite{Zhang2025PhaseTrackedEncircling}. Yet state transfer is not by itself a cooling advantage, and chiral conversion can occur even without EP encircling \cite{Hassan2017WithoutEncircling,Nasari2022ObservationWithoutEncircling}.  These advances establish that EP structure can redirect dynamics, but they leave open a more demanding question: \emph{can encircling itself improve finite-time cooling when the enclosing and non-enclosing controls are given the same physical resources?}

Here we implement that comparison in the three-mode system sketched in Fig.~\ref{Schematic}. A strongly damped main cavity provides the optical loss channel. It is coherently coupled to an auxiliary cavity whose linewidth is exactly zero, so the auxiliary mode can store and exchange excitation but cannot remove it directly. Together with the mechanics, the two cavities produce three hybrid branches and two EPs that connect different branch pairs. We fix one periodic power modulation and one periodic detuning waveform, then change only their relative phase. One phase gives a loop that encloses neither EP; another gives a loop that encloses both. Because the second detuning is only a cyclic translation of the first, the duration, endpoint, mean, range, integrated effort, maximum slew rate, and maximum acceleration are unchanged. The topology therefore changes without giving either protocol a different waveform shape or a larger local control budget.

The two-EP phase leaves less than half the final mechanical occupation of the non-enclosing phase. We then test whether this separation survives the complete allowed phase family, substantial changes of waveform shape, and a finite neighborhood of nearby smooth controls. After establishing this robustness, we resolve how excitation moves through the lossless auxiliary mode and the lossy main cavity. Finally, we keep the controls fixed and restore the rapidly oscillating heating processes that were neglected when defining the EP spectrum. This sequence separates the controlled topology comparison, its robustness, its physical mechanism, and the validation of the dynamical approximation.
\begin{figure}[t]
\centering
\includegraphics[width=1\linewidth]{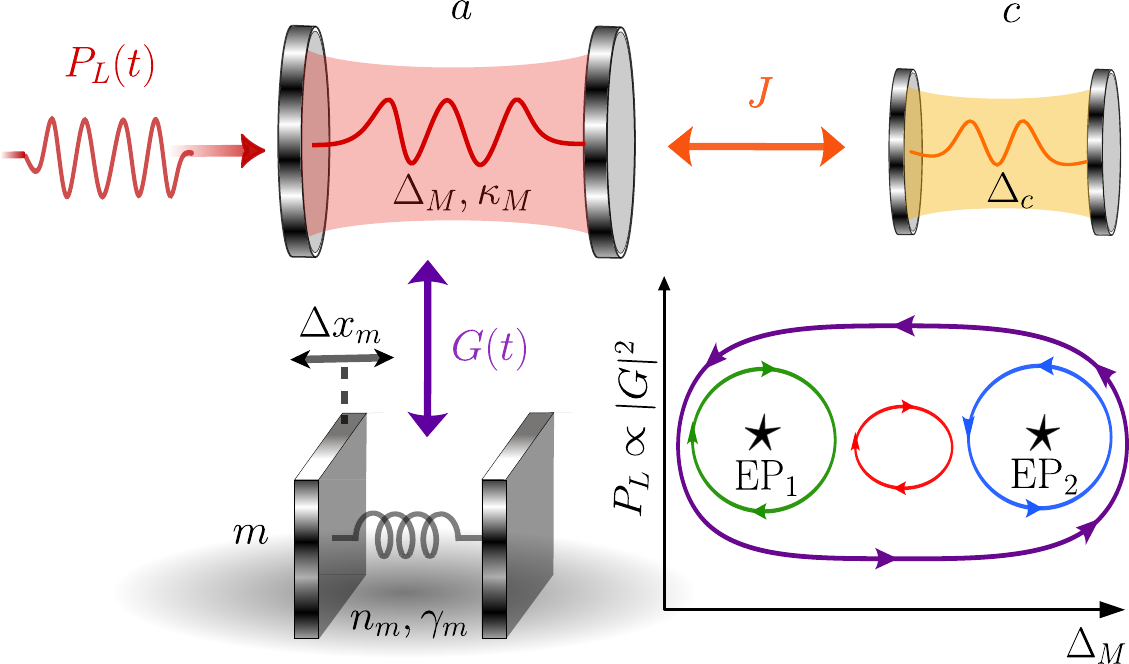}
\caption{Lossless spectral-routing architecture. A driven main cavity couples the mechanical resonator to the optical loss channel and is coherently coupled to an auxiliary cavity. The auxiliary linewidth is exactly zero, so this mode cannot cool by direct dissipation. Its role is to add a third coherent hybrid branch. Two second-order exceptional points connect different pairs of the three branches in the control plane; a loop around both can therefore compose the two pairwise exchanges into a three-branch cycle.\justifying}
\label{fig:schematic}
\end{figure}

Figure~\ref{fig:schematic} summarizes the separation of physical roles used throughout the paper. The main cavity couples the mechanics to an irreversible optical outlet. The auxiliary cavity is connected only coherently and has no loss channel of its own. Its purpose is instead spectral: it adds a third hybrid branch, allowing two different pairwise exceptional-point connections to coexist. This distinction will let us test spectral routing without adding a second dissipative resource.

\textit{Model}.--
We now define the dynamics used for both the spectral analysis and the cooling calculation. The main optical cavity $a$ couples to the mechanical mode $m$ through radiation pressure and to the auxiliary cavity $c$ through coherent photon hopping. After displacing the driven main-cavity field and linearizing the radiation-pressure interaction, red-sideband driving makes the excitation-exchange terms resonant. We first retain these terms within the rotating-wave approximation (RWA), which gives a time-local three-branch spectrum and therefore a precise exceptional-point geometry. The counter-rotating pair-creation terms are restored later without changing the controls. The RWA Hamiltonian is
\begin{equation}
\frac{H_{\rm RWA}}{\hbar}
=
-\Delta_M a^\dagger a
-\Delta_c c^\dagger c
+G a^\dagger m+G^\ast a m^\dagger
+J a^\dagger c+J^\ast a c^\dagger.
\label{eq:main_rwa_hamiltonian}
\end{equation}
Here $\Delta_M$ is the controlled main-cavity detuning, $\Delta_c$ is the fixed auxiliary-cavity detuning, $G$ is the drive-enhanced optomechanical coupling, and $J$ is the intercavity coupling. We use $P_L\equiv|G|^2$ as the power-like control coordinate. It measures the squared enhanced coupling and is not a separately calibrated laboratory laser power.

We measure frequencies in units of the main-cavity linewidth $\kappa_M$ and use the normalized time $\tau=\kappa_M t$. We also define $\delta_M=\Delta_M/\kappa_M$, $\delta_c=\Delta_c/\kappa_M$, $g=G/\kappa_M$, $j=J/\kappa_M$, and $\widetilde\gamma_m=\gamma_m/\kappa_M$. For the annihilation-operator vector $\bm v=(a,m,c)^T$, the full three-mode RWA drift is
\begin{equation}
A_{\rm full}(\tau)
=
\begin{pmatrix}
 i\delta_M(\tau)-1/2 & -ig(\tau) & -ij\\
 -ig^\ast(\tau) & -\widetilde\gamma_m/2 & 0\\
 -ij^\ast & 0 & i\delta_c
\end{pmatrix}.
\label{eq:main_full_drift}
\end{equation}
The main cavity is damped, the mechanics is weakly damped, and the auxiliary cavity has no damping term. Thus $\kappa_c=0$ exactly. The auxiliary mode is a coherent degree of freedom, not a second optical reservoir.

Unless stated otherwise, throughout the manuscript we use \(\delta_c=0.1\), \(j=0.1\), \(\widetilde{\kappa}_c=5\times10^{-3}\), and \(\widetilde{\gamma}_m=10^{-4}\), together with \(n_M^{\rm th}=n_c^{\rm th}=0\), \(n_m^{\rm th}=100\), and the initial occupations \(N(0)={\rm diag}(0,100,0)\). For the full Bogoliubov validation we additionally use \(\Omega_m=\omega_m/\kappa_M=10\). The main-cavity bath is in vacuum, while the lossless auxiliary mode has no bath. The normally ordered covariance matrix $N_{ij}(\tau)=\langle v_i^\dagger(\tau)v_j(\tau)\rangle$ obeys
\begin{equation}
\frac{\dd N}{\dd\tau}
=
A_{\rm full}^{\ast}N
+NA_{\rm full}^{T}
+D_{\rm th},
\quad
D_{\rm th}
=
{\rm diag}\!\left(0,\widetilde\gamma_m n_m^{\rm th},0\right).
\label{eq:main_covariance}
\end{equation}
Our cooling observable is the mechanical occupation $n_m(\tau)=N_{mm}(\tau)$. Equation~\eqref{eq:main_full_drift} therefore determines the instantaneous hybrid spectrum, while Eq.~\eqref{eq:main_covariance} determines how the physical state actually evolves through that spectrum.

\textit{Two exceptional points and a same-waveform topology toggle}.--
With the physical model fixed, the next question is whether its spectrum contains the connectivity needed to involve all three modes. A single second-order EP connects only two sheets. Here the three-mode drift has two accessible second-order EPs that connect different pairs,
\begin{equation}
\begin{aligned}
{\rm EP}_1:&\quad
(\Delta_M,P_L)=(-3.52,\,5.38),\\
{\rm EP}_2:&\quad
(\Delta_M,P_L)=(0.14,\,5.37).
\end{aligned}
\label{eq:main_lossless_eps}
\end{equation}
At the common start and end point of the control cycle, we label the continuously tracked branches by their dominant bare-mode character: branch $0$ is mainly main-cavity-like, branch $1$ is mainly mechanical-like, and branch $2$ is mainly auxiliary-like. At ${\rm EP}_1$, branches $1$ and $2$ merge; at ${\rm EP}_2$, branches $0$ and $1$ merge. A loop around both points can therefore compose two different pair exchanges into a three-branch cycle. The EPs are found from the full three-mode drift. We verify both eigenvalue and eigenvector coalescence, and we show that the same defects are embedded in the full Lindblad generator, or Liouvillian, in Note~\ref{suppnote:model} of the SM \cite{SuppMat}.

The existence of two EPs does not by itself show that they improve cooling. We still need two controls whose local resources are matched while their global routes differ. We therefore fix the coupling-power waveform for the entire comparison,
\begin{equation}
p_L(\tau)
=
p_c+r_p\cos\!\left(\frac{2\pi\tau}{\mathcal T}\right),
\quad
p_c=0.05,
\quad
r_p=0.01,
\label{eq:main_fixed_power}
\end{equation}
where $p_L=P_L/\kappa_M^2=|g|^2$. We also fix one three-harmonic detuning shape,
\begin{equation}
\delta_0(\tau)
=
\bar\delta+f\!\left(\frac{2\pi\tau}{\mathcal T}\right),
\quad
\bar\delta=-0.25,
\label{eq:main_base_detuning}
\end{equation}
with the coefficients of $f$ given in Note~\ref{suppnote:phase_toggle} of the SM \cite{SuppMat}. The non-enclosing protocol uses this waveform as written. The two-EP protocol uses the same periodic waveform shifted by the phase $\theta_\ast=4.2$ relative to the unchanged power modulation.

This shift is chosen so that the translated waveform starts and ends at exactly the same detuning as the original waveform but encircling the two EPs. Because it is a translation over a full period, it also preserves the detuning mean, range, integrated derivative effort, maximum slew rate, and maximum acceleration. We label a closed path by $(w_1,w_2)$, where $w_\mu$ is its signed winding number around ${\rm EP}_\mu$. The unshifted loop, denoted $\controlclass_0$, has $(w_1,w_2)=(0,0)$; the shifted loop, denoted $\controlclass_{12}$, has $(w_1,w_2)=(1,1)$. Thus the comparison changes only the relative timing of the same local waveforms and, with it, the global EP connectivity.
\begin{figure}[t]
\centering
\includegraphics[width=1\linewidth]{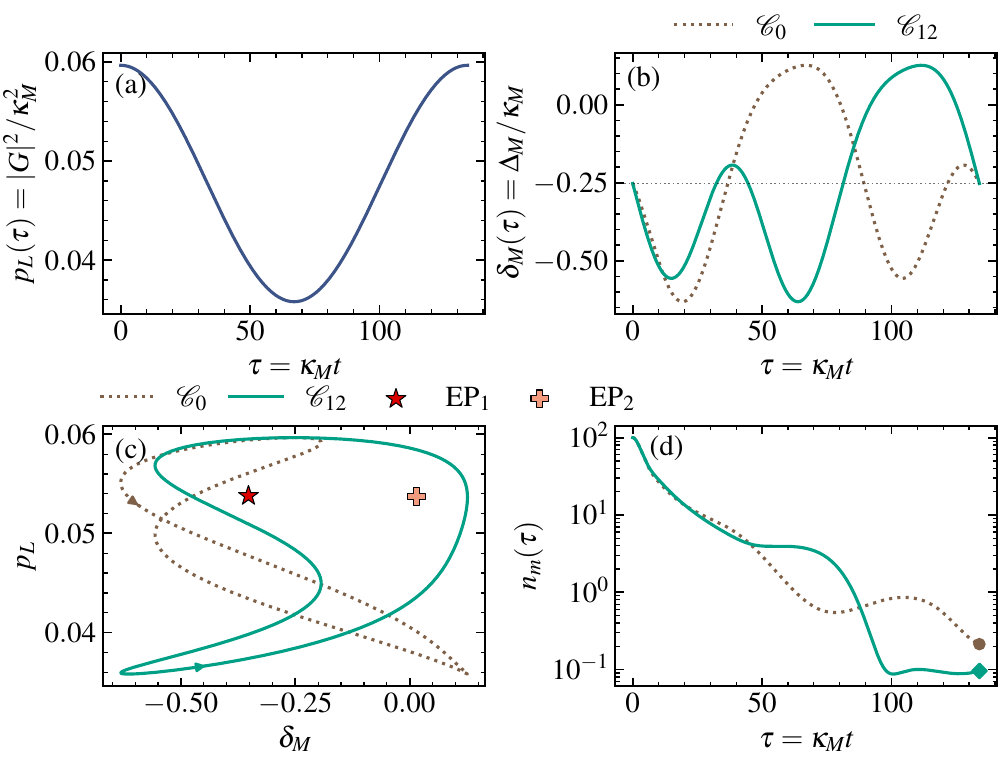}
\caption{Same-waveform topology toggle. (a) The coupling-power waveform is identical in the two protocols. (b) The main-cavity detuning has the same periodic shape in both cases and differs only by a cyclic phase shift. (c) The shift changes the closed control path from the non-enclosing class $\controlclass_0$ to the two-EP class $\controlclass_{12}$. The star and plus markers identify the two exceptional points. (d) Mechanical occupation during the cycle. The endpoint values are interval certified.\justifying}
\label{fig:phase_toggle}
\end{figure}

Figure~\figpanel{fig:phase_toggle}{a} makes the common coupling resource explicit: the two power traces coincide point by point. Figure~\figpanel{fig:phase_toggle}{b} shows the only local change, namely the cyclic translation of one fixed detuning shape. In the control plane, Fig.~\figpanel{fig:phase_toggle}{c} shows why that translation matters globally: the brown path returns without enclosing either EP, whereas the green path winds around both. The dynamical consequence appears in Fig.~\figpanel{fig:phase_toggle}{d}. The two occupations are not ordered at every instant and cross during the evolution, but they separate strongly near the end of the cycle.

A rigorous interval propagation bounds the completed-cycle occupations as
\begin{equation}
\begin{aligned}
n_m^{\controlclass_0}(\mathcal T)
&\in[0.2128679579,\,0.2128686346],\\
n_m^{\controlclass_{12}}(\mathcal T)
&\in[0.0956793480,\,0.0956798219].
\end{aligned}
\label{eq:main_phase_toggle_bounds}
\end{equation}
The two-EP loop therefore reduces the final mechanical occupation by at least $55.0520\%$ relative to the matched non-enclosing loop. Since the curves cross, this is not a uniform increase of an instantaneous cooling rate. It is a completed-cycle result produced by a different time ordering of hybridization, coherent storage, and exposure to the same main-cavity loss. Note~\ref{suppnote:phase_toggle} of the SM \cite{SuppMat} gives the exact resource identities, topology certificate, interval propagation, and independent covariance cross-check.

\textit{Robustness in phase and waveform space}.--
The matched pair in Fig.~\ref{fig:phase_toggle} removes the optimization ambiguity, but one concern remains: the selected phase could be an isolated favorable point. We address this first without changing the underlying waveform shape. Let $x=2\pi\tau/\mathcal T$ and let $f(x)$ be the three-harmonic function above. We define the endpoint-anchored family
\begin{equation}
F_\theta(x)
=
f(x+\theta)-f(\theta)\cos x,
\quad
\delta_\theta(x)
=
\bar\delta+s_\theta F_\theta(x),
\label{eq:main_anchored_family}
\end{equation}
where the positive scale $s_\theta$ restores the same integrated detuning effort. The subtraction fixes the common endpoint and mean. At the two phase-toggle values, the correction vanishes and the family reproduces the controls of Fig.~\ref{fig:phase_toggle} exactly.

\begin{figure}[t]
\centering
\includegraphics[width=1\linewidth]{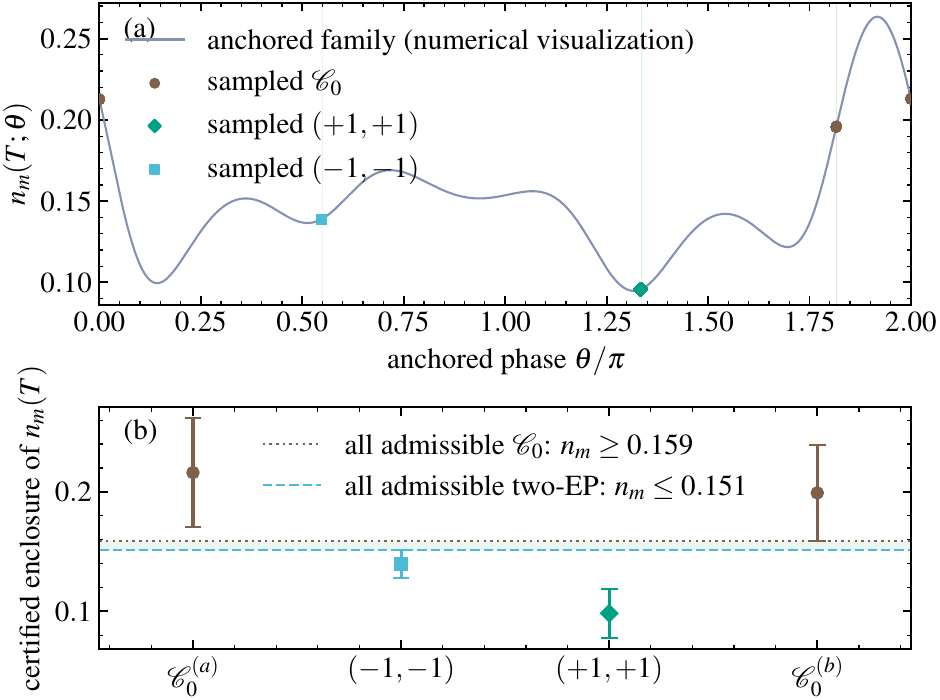}
\caption{Complete endpoint-anchored phase family. (a) Numerical visualization of the final occupation around the full phase circle, with representative points from the four interval-certified candidate sectors. (b) Conservative occupation enclosures for the complete phase hulls. The horizontal bounds show that every admissible two-EP member lies below every admissible non-enclosing member within this family.\justifying}
\label{fig:phase_family}
\end{figure}

We then cover the full phase circle with interval arithmetic rather than judging admissibility from a dense numerical scan. We call a member admissible only if it satisfies the common detuning-extremum bands, slew and acceleration limits, EP-clearance requirement, and topology condition. Every fully admissible member lies in one of four small sectors: two non-enclosing sectors and two sectors that wind around both EPs with opposite orientations. No admissible single-EP sector remains in this anchored family. Figure~\figpanel{fig:phase_family}{a} shows the numerical occupation along the full family and marks representative points from the four sectors. This curve is only a visual guide. The proof is summarized in Fig.~\figpanel{fig:phase_family}{b}, where each error bar encloses the complete conservative phase hull, including some points that may fail an extrema condition. Even with this deliberately loose enclosure, the two two-EP hulls lie below the non-enclosing lower bound.

Quantitatively, the family-wide bounds are
\begin{align}\nonumber
\sup_{\substack{\theta\ {\rm admissible}\\ |w_1|=|w_2|=1}}
n_m(\mathcal T;\theta)
&\le0.1512866265
<0.1589328792\\
&\le
\inf_{\substack{\theta\ {\rm admissible}\\ (w_1,w_2)=(0,0)}}
n_m(\mathcal T;\theta).
\label{eq:main_levelB_all_to_all}
\end{align}
Thus, within the entire admissible endpoint-anchored family, every two-EP control finishes colder than every non-enclosing control. Note~\ref{suppnote:phase_family} of the SM \cite{SuppMat} gives the exhaustive cover, the four conservative phase hulls, and the interval dynamics used to prove this ordering.

\begin{figure}[t]
\centering
\includegraphics[width=1\linewidth]{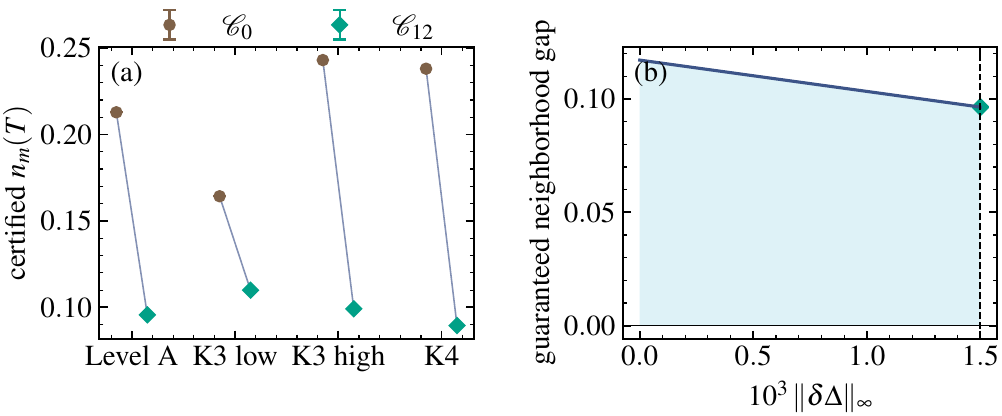}
\caption{Robustness beyond one waveform. (a) Interval-certified final occupations for the phase-toggle reference and three predeclared spectral redistributions. Each vertical pair compares a non-enclosing and a two-EP control under the same resources. (b) Guaranteed separation between the two finite topology neighborhoods as the detuning perturbation radius is increased to its full certified value.\justifying}
\label{fig:waveform_robustness}
\end{figure}

The anchored family still changes only the phase of one basic waveform. We therefore next ask whether the hierarchy survives when the waveform itself is changed. The alternative Fourier shapes are fixed before the cooling objective is evaluated: one transfers ten percentage points of derivative effort from the second harmonic to the first, one transfers the same amount from the second to the third, and one introduces a new fourth harmonic carrying five percent of the total effort. For each shape, the comparison phases are chosen using only the control geometry, topology, minimum EP distance, and derivative margins. Figure~\figpanel{fig:waveform_robustness}{a} shows the certified endpoints. All three modified shapes preserve the ordering, with guaranteed reductions of at least $33.0769\%$, $59.1932\%$, and $62.4103\%$.

A finite set of shapes still does not prove that the result survives continuously small deformations. We therefore certify a neighborhood of arbitrary smooth perturbations around the phase-toggle pair. Writing the detuning perturbation away from either reference control as $\delta\Delta$, we restrict to controls that preserve the endpoint, mean, and effort exactly and satisfy
\begin{equation}
\|\delta\Delta\|_\infty\le1.5\times10^{-3},
\quad
\|\delta\dot\Delta\|_\infty\le0.20,
\quad
\|\delta\ddot\Delta\|_\infty\le0.20.
\label{eq:main_open_neighborhood}
\end{equation}
Every control in the stated $\controlclass_0$ neighborhood satisfies $n_m(\mathcal T)\ge0.1993737215$, whereas every control in the stated $\controlclass_{12}$ neighborhood satisfies $n_m(\mathcal T)\le0.1029231965$. Figure~\figpanel{fig:waveform_robustness}{b} follows the guaranteed separation as the detuning radius grows from zero to the full certified value. The gap stays strictly positive throughout. An independent periodic cubic spline gives the same ordering numerically without using the Fourier representation.

The shape-selection rule, the corresponding control paths and detuning waveforms, the independent spline cross-check, and the finite-neighborhood proof are given in Note~\ref{suppnote:waveform_robustness} of the SM \cite{SuppMat}.

\textit{How the lossless braid improves cooling}.--
The robustness results establish that the separation is not a single lucky phase or a single Fourier shape. We now ask what the two-EP route changes physically. The exactly lossless auxiliary cavity makes this question unusually clean. Since $\kappa_c=0$, no excitation can leave through the auxiliary mode. It can only be stored there, exchanged coherently with the main cavity, or redistributed through hybridization. All irreversible optical removal occurs through the main cavity.

Because the covariance equation is linear, the final occupation can be split exactly into two parts: excitation propagated from the initial state and excitation injected by the mechanical bath during the cycle. This source decomposition tells us which part is most affected. In the RWA, the $\controlclass_0$ endpoint contains $0.1329699$ from the initial covariance and $0.0798984$ from mechanical-bath injection. For $\controlclass_{12}$ the corresponding contributions are $0.0288494$ and $0.0668302$. Hence $88.85\%$ of the final $\controlclass_0-\controlclass_{12}$ separation comes from more complete removal of excitation that was already present in the mechanics at the beginning of the cycle. The braid also reduces the bath-generated contribution, but that is the smaller part of the main gap.

\begin{figure}[t]
\centering
\includegraphics[width=1\linewidth]{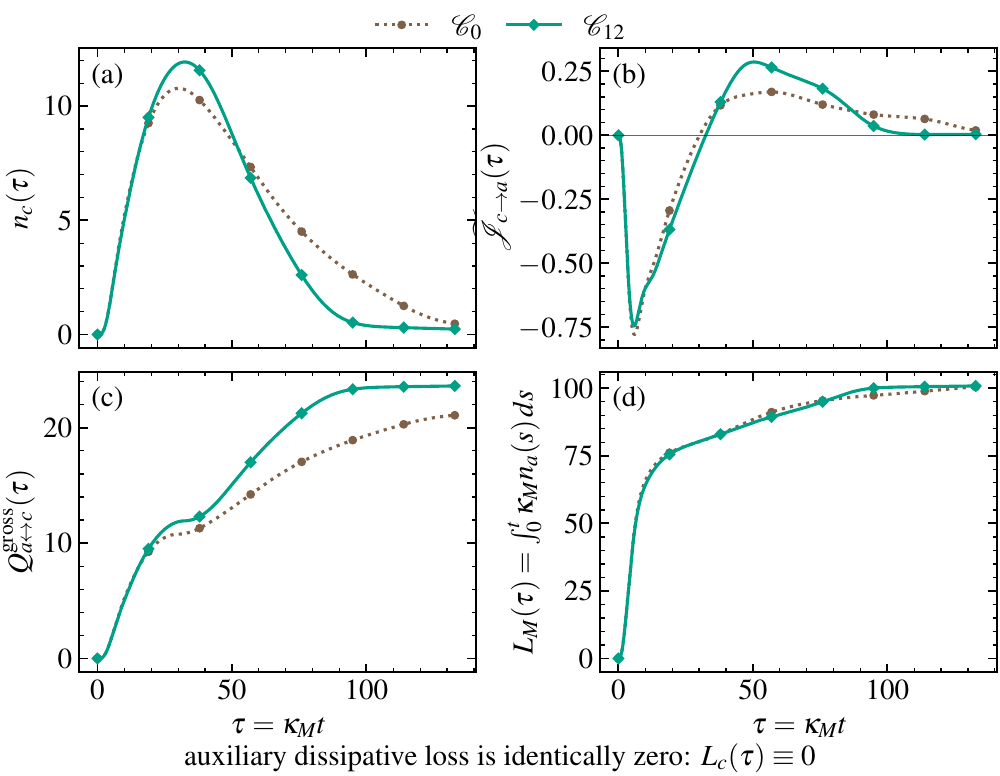}
\caption{Coherent role of the lossless auxiliary cavity for the phase-toggle pair. (a) Auxiliary occupation. (b) Coherent auxiliary-to-main-cavity current; the sign change shows repeated storage and return. (c) Cumulative gross bidirectional exchange between the two optical modes. (d) Cumulative main-cavity loss. The auxiliary dissipative loss is identically zero throughout the cycle.\justifying}
\label{fig:mechanism}
\end{figure}

Figure~\ref{fig:mechanism} resolves the coherent and dissipative channels. Figure~\figpanel{fig:mechanism}{a} shows that the auxiliary mode becomes strongly populated during both protocols, reaching an even larger peak for the two-EP loop. Figure~\figpanel{fig:mechanism}{b} shows the coherent current between the auxiliary and main cavities; its sign changes because excitation is repeatedly stored and returned rather than flowing irreversibly in one direction. The accumulated magnitude of this exchange is shown in Fig.~\figpanel{fig:mechanism}{c}: the completed gross exchange is about $21.11$ for $\controlclass_0$ and $23.63$ for $\controlclass_{12}$. Yet the auxiliary dissipative loss is identically zero. Figure~\figpanel{fig:mechanism}{d} shows the only optical sink, the cumulative main-cavity loss. Its final values, about $100.53$ and $100.89$, are very close.

These panels rule out a simple ``more dissipation'' explanation. The two controls lose almost the same total amount through the main cavity, but they leave very different mechanical populations at the final time. The relevant difference is therefore the timing of coherent redistribution relative to loss. The two-EP connectivity changes when mechanical, auxiliary, and main-cavity character appear along the connected spectral route, and the exact dynamics determines how much excitation uses each part of that route. Note~\ref{suppnote:mechanism} of the SM \cite{SuppMat} derives the excitation-balance equations and shows the corresponding tracked branch compositions.

\textit{Fixed-control beyond-RWA validation}.--
One modeling approximation remains. The EP geometry above is defined by the red-sideband RWA drift. The full linearized optomechanical interaction also contains rapidly oscillating counter-rotating terms that create or annihilate a photon--phonon pair; these are the Stokes processes omitted by the RWA and they can add heating. We therefore do not reoptimize the controls. Instead, we keep the phase-toggle pair and representative robustness controls fixed and enlarge only the dynamical model. Setting the counter-rotating block to zero in the doubled Bogoliubov--de Gennes (BdG) calculation reproduces the RWA endpoints with a maximum discrepancy of $2.41\times10^{-12}$, which provides an internal check of the implementation.

Restoring the counter-rotating terms changes the phase-toggle endpoint from $0.2128683$ to $0.2455145$ for $\controlclass_0$ and from $0.0956796$ to $0.0973702$ for $\controlclass_{12}$. The two-EP control therefore remains $60.3404\%$ below the non-enclosing control in the full Bogoliubov dynamics. Representative controls from the complete phase family and the waveform-shape test retain reductions of $34.6173\%$ and $64.5276\%$, respectively. Figure~\ref{fig:supp_bdg_validation} and Note~\ref{suppnote:bdg} of the SM \cite{SuppMat} give the full time traces, endpoint comparison, and source decomposition. The finite-neighborhood theorem itself remains an RWA theorem: the Bogoliubov calculation validates its center pair but does not certify the entire neighborhood.

\textit{Summary and discussion}.--
Our central result is a controlled topology toggle: a cyclic phase shift changes only the relative timing between one fixed detuning waveform and one fixed power modulation, turning a non-enclosing path into a loop around both EPs while preserving the local control resources and reducing the final mechanical occupation by more than half. The advantage persists across the complete anchored phase family, substantial waveform-shape changes, an independent spline representation, a finite smooth neighborhood, and fixed-control Bogoliubov dynamics with counter-rotating heating restored. In the lossless auxiliary limit, the mechanism is especially clear: the auxiliary cavity does not cool by dissipation, but instead provides coherent storage and the extra spectral sheet needed for the two-EP braid, while the main cavity remains the only optical sink; the braid changes when excitation-bearing hybrid character is exposed to that loss. We therefore establish a causal and robust finite-time cooling advantage from access to the two-EP connectivity under matched resources, without claiming that topology alone fixes the cooling rate or globally orders all smooth controls.
\acknowledgments{\emph{Acknowledgments}---BA acknowledges support from IRA Program (project no. FENG.02.01-IP.05-0006/23) financed by the FENG program 2021--2027, Priority FENG.02, Measure FENG.02.01., with the support of the FNP.}

\emph{Data and code availability}.---
The numerical code, interval certificates, and plotting scripts used to generate the results in this work will be made available in the project repository. The release will contain the lossless phase-toggle certificate, the complete anchored-family certificate, the waveform-robustness certificates, and the fixed-control Bogoliubov and mechanism calculations used here.

\bibliography{References}

\clearpage
\onecolumngrid

\setcounter{section}{0}
\setcounter{subsection}{0}
\setcounter{equation}{0}
\setcounter{figure}{0}
\setcounter{table}{0}

\renewcommand{\thesection}{\arabic{section}}
\renewcommand{\thesubsection}{\Alph{subsection}}
\renewcommand{\theequation}{S\arabic{equation}}
\renewcommand{\thefigure}{S\arabic{figure}}
\renewcommand{\thetable}{S\arabic{table}}

\newcommand{\suppnote}[1]{%
    \refstepcounter{section}%
    \setcounter{subsection}{0}%
    \section*{Supplementary Note \arabic{section}. #1}%
    \addcontentsline{toc}{section}{Supplementary Note \arabic{section}. #1}%
}

\newcommand{\suppsubsection}[1]{%
    \refstepcounter{subsection}%
    \subsection*{\Alph{subsection}. #1}%
    \addcontentsline{toc}{subsection}{\Alph{subsection}. #1}%
}

\section*{Supplemental Material for ``Finite-Time Optomechanical Cooling by Multi-Exceptional-Point Braiding''}

The Supplemental Material follows the logic of the main text and supplies the derivations and certificates behind each step. Note~\ref{suppnote:model} derives the lossless three-mode model, establishes the two exceptional points, and embeds their defects in the full Lindblad generator (Liouvillian). Note~\ref{suppnote:phase_toggle} constructs the same-waveform phase toggle and certifies its matched resources and endpoint occupations. Note~\ref{suppnote:phase_family} extends the comparison over the complete endpoint-anchored phase family. Note~\ref{suppnote:waveform_robustness} tests changes of waveform shape and proves a finite smooth neighborhood. Note~\ref{suppnote:bdg} restores the counter-rotating interaction at fixed controls. Note~\ref{suppnote:mechanism} then resolves the coherent exchange, loss balance, and tracked spectral composition.

\suppnote{Lossless three-mode model and exceptional-point structure}
\label{suppnote:model}

This note establishes the physical model before any topology or cooling comparison is made. We first derive the linearized lossless three-mode dynamics, then locate the two exceptional points directly in the full drift, and finally show why these drift singularities are also singularities of the full linear Lindblad generator. The covariance equation and a non-expansiveness identity needed for the later certificates are given at the end.

\suppsubsection{Linearized rotating-wave model}
\label{suppsec:rwa_model}

We begin from the standard radiation-pressure coupling so that the origin of every term in the drift matrix is explicit. Before linearization, the interaction of the main cavity with the mechanical mode is
\begin{equation}
H_{\rm rp}
=
-\hbar g_0 A^\dagger A(m+m^\dagger),
\label{eq:supp_rp}
\end{equation}
where $A$ is the full main-cavity field and $g_0$ is the single-photon optomechanical coupling. Under coherent driving we write $A=\alpha+a$, with $\alpha$ the classical intracavity amplitude and $a$ the fluctuation operator. Neglecting the nonlinear fluctuation term gives
\begin{equation}
H_{\rm lin}
=
-\hbar(Ga^\dagger+G^\ast a)(m+m^\dagger),
\quad
G=g_0\alpha.
\label{eq:supp_linearized_interaction}
\end{equation}
For red-sideband driving, the rotating-wave approximation (RWA) keeps the beam-splitter terms and gives Eq.~\eqref{eq:main_rwa_hamiltonian} of the main text.

Including the main-cavity and mechanical reservoirs while setting the auxiliary linewidth exactly to zero gives
\begin{equation}
\begin{aligned}
\dot a
&=
\left(i\Delta_M-\frac{\kappa_M}{2}\right)a-iGm-iJc+\xi_a,\\
\dot m
&=
-\frac{\gamma_m}{2}m-iG^\ast a+\xi_m,\\
\dot c
&=
i\Delta_c c-iJ^\ast a.
\end{aligned}
\label{eq:supp_lossless_langevin}
\end{equation}
There is no auxiliary input-noise operator because $\kappa_c=0$. The physical drift matrix is therefore
\begin{equation}
A_{\rm phys}
=
\begin{pmatrix}
 i\Delta_M-\kappa_M/2 & -iG & -iJ\\
 -iG^\ast & -\gamma_m/2 & 0\\
 -iJ^\ast & 0 & i\Delta_c
\end{pmatrix}.
\label{eq:supp_physical_drift}
\end{equation}
Dividing by $\kappa_M$ gives Eq.~\eqref{eq:main_full_drift}.

The parameters used throughout the lossless calculations are
\begin{equation}
\kappa_M=10,
\quad
\gamma_m=10^{-3},
\quad
J=1,
\quad
\Delta_c=1,
\quad
\kappa_c=0,
\label{eq:supp_parameters}
\end{equation}
with mechanical frequency $\omega_m=100$ for the beyond-RWA calculation. The loop duration is $T=13.4$, so the normalized duration is $\mathcal T=134$. The mechanical bath and the initial mechanical state both have occupation $100$, while the optical baths and optical modes start in vacuum.

\suppsubsection{Exact lossless exceptional points}
\label{suppsec:lossless_eps}

We now ask where two instantaneous hybrid modes actually become defective. For fixed $\Delta_c$ and $J$, let
\begin{equation}
p(\lambda;\Delta_M,P_L)
=
\det\!\left[\lambda I_3-A_{\rm phys}(\Delta_M,P_L)\right],
\quad
P_L=|G|^2.
\label{eq:supp_characteristic_polynomial}
\end{equation}
Writing $p(\lambda)=\lambda^3+\alpha\lambda^2+\beta\lambda+\gamma$, the cubic discriminant is
\begin{equation}
\mathscr D
=
\alpha^2\beta^2-4\beta^3-4\alpha^3\gamma-27\gamma^2+18\alpha\beta\gamma
=
\prod_{i<j}(\lambda_i-\lambda_j)^2.
\label{eq:supp_discriminant}
\end{equation}
A zero of $\mathscr D$ gives an eigenvalue degeneracy. We then verify that the degenerate eigenvalue has only one independent right eigenvector, which distinguishes an exceptional point from an ordinary diagonalizable degeneracy. Interval root isolation gives exactly two accessible second-order exceptional points in the control region,
\begin{table}[t]
\centering
\begin{ruledtabular}
\begin{tabular}{lcccc}
 & $\Delta_M$ & $P_L$ & $\delta_M$ & $p_L$\\
\hline
${\rm EP}_1$ & $-3.5244934644$ & $5.3798272216$ & $-0.3524493464$ & $0.0537982722$\\
${\rm EP}_2$ & $0.1426164653$ & $5.3734870080$ & $0.0142616465$ & $0.0537348701$
\end{tabular}
\end{ruledtabular}
\caption{Exceptional-point coordinates of the lossless three-mode drift. The normalized coordinates are $\delta_M=\Delta_M/\kappa_M$ and $p_L=P_L/\kappa_M^2$.\justifying}
\label{tab:supp_lossless_eps}
\end{table}
At the common start and end point we order the branches by dominant bare-mode character: branch $0$ is main-cavity-like, branch $1$ is mechanical-like, and branch $2$ is auxiliary-like. At ${\rm EP}_1$ branches $1$ and $2$ coalesce; at ${\rm EP}_2$ branches $0$ and $1$ coalesce.

\begin{figure}[t]
\centering
\includegraphics[width=0.70\linewidth]{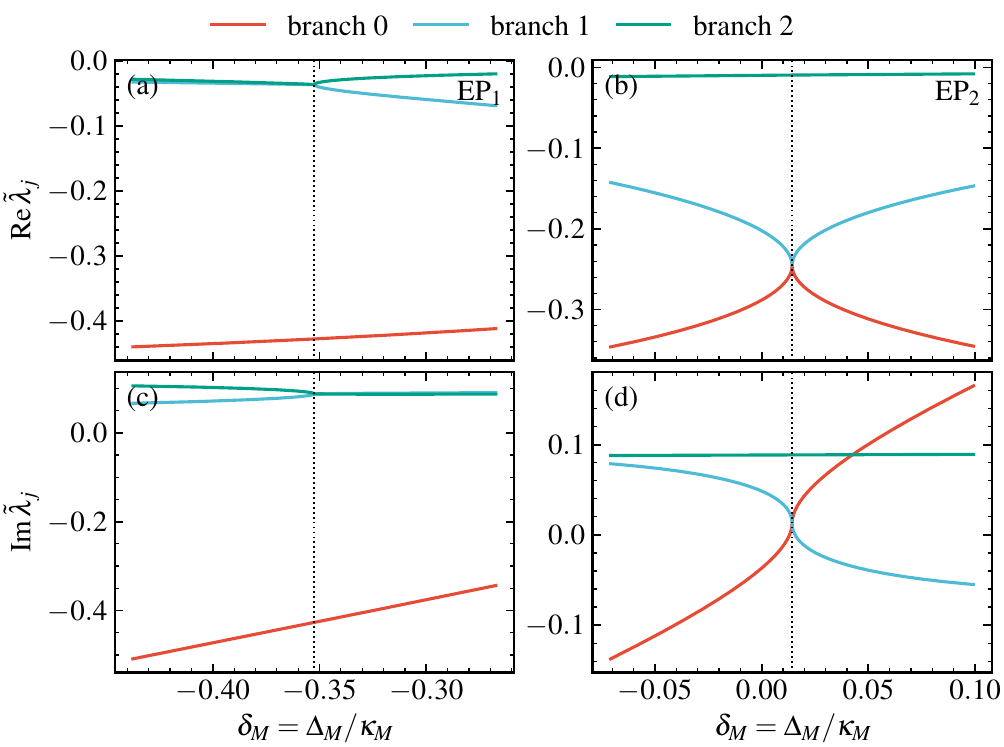}
\caption{Spectral cuts through the two lossless exceptional points. The real and imaginary parts of the three continuously tracked drift eigenvalues are shown as the main-cavity detuning crosses each degeneracy at fixed coupling power. The two coalescing sheets merge in both parts of the eigenvalue while the third sheet remains separated.\justifying}
\label{fig:supp_lossless_ep_cuts}
\end{figure}

Figure~\figpanel{fig:supp_lossless_ep_cuts}{a} and Fig.~\figpanel{fig:supp_lossless_ep_cuts}{c} show the real and imaginary parts of the spectrum through ${\rm EP}_1$. Branches $1$ and $2$ meet simultaneously in both parts, while branch $0$ stays separated. Figure~\figpanel{fig:supp_lossless_ep_cuts}{b} and Fig.~\figpanel{fig:supp_lossless_ep_cuts}{d} show the complementary event at ${\rm EP}_2$, where branches $0$ and $1$ merge while branch $2$ remains distinct. The two points therefore connect different branch pairs; this difference is what makes a composed three-branch cycle possible.

The exceptional points are not artifacts of a first-moment surrogate. At fixed parameters, the adjoint Lindblad generator (Liouvillian) acts on the exact invariant operator space ${\cal V}_1={\rm span}\{a,m,c\}$ as
\begin{equation}
\mathcal L^\dagger\bm v=A_{\rm phys}\bm v.
\label{eq:supp_liouvillian_embedding}
\end{equation}
A Jordan defect of $A_{\rm phys}$ is therefore a Jordan defect of the restriction of the full adjoint Liouvillian to ${\cal V}_1$. A diagonalizable full operator cannot have a non-diagonalizable restriction to an invariant subspace. Hence the two drift exceptional points are also Liouvillian exceptional points of the complete linear RWA Lindblad dynamics \cite{PerinaJr2022QuantumLiouvillian,Gaidash2025LindbladBosonic}.

\begin{figure}[t]
\centering
\includegraphics[width=0.70\linewidth]{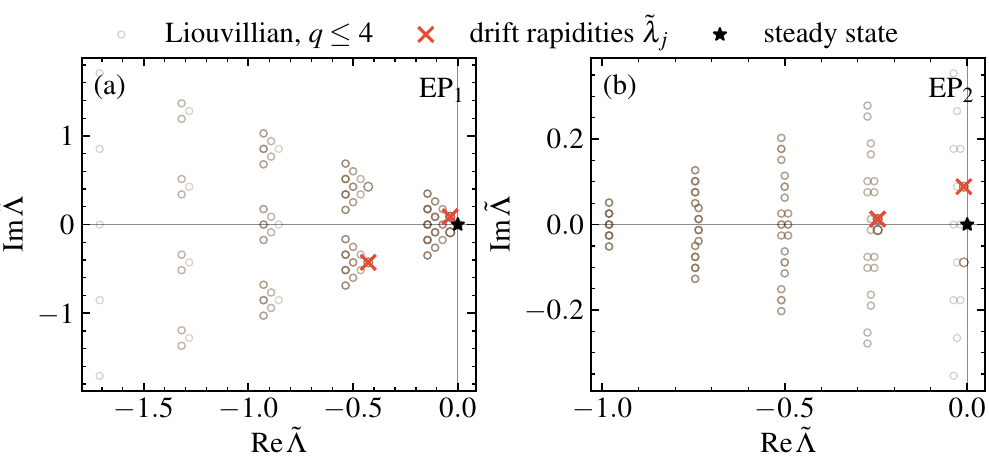}
\caption{Embedding of the lossless drift eigenvalues in the exact bosonic Liouvillian spectral ladder at the two exceptional points. The displayed finite-order ladder is generated from the exact elementary rapidities; it is not a Hilbert-space truncation.\justifying}
\label{fig:supp_liouvillian_ladder}
\end{figure}

For the passive quadratic RWA model, the full Liouvillian spectrum is generated from the drift eigenvalues $\lambda_j$, often called rapidities, as
\begin{equation}
\Lambda_{\bm n,\bm m}
=
\sum_{j=0}^{2}n_j\lambda_j
+
\sum_{j=0}^{2}m_j\lambda_j^\ast,
\quad
n_j,m_j\in\mathbb N_0.
\label{eq:supp_liouvillian_ladder}
\end{equation}
The first-order sector contains the three drift eigenvalues and their conjugates. Figure~\ref{fig:supp_liouvillian_ladder} shows this embedding explicitly.

Figure~\figpanel{fig:supp_liouvillian_ladder}{a} and Fig.~\figpanel{fig:supp_liouvillian_ladder}{b} display this construction at ${\rm EP}_1$ and ${\rm EP}_2$, respectively. In each panel the elementary drift eigenvalues sit exactly in the first-order Liouvillian sector, while the higher-order points are sums of the same rapidities. The plot is therefore a spectral embedding check, not a finite Hilbert-space approximation.

\suppsubsection{Covariance evolution and non-expansiveness}
\label{suppsec:covariance_nonexpansive}

Having established the spectral structure, we now state the dynamical property used in the interval proofs. The normally ordered covariance obeys Eq.~\eqref{eq:main_covariance} \cite{Serafini2017QuantumCV}. In physical units the homogeneous drift satisfies the exact identity
\begin{equation}
A_{\rm phys}+A_{\rm phys}^\dagger
=
{\rm diag}(-10,-0.001,0)
\preceq0.
\label{eq:supp_nonexpansive_identity}
\end{equation}
Therefore the homogeneous propagator is non-expansive in the Euclidean norm. The zero in the auxiliary entry should not be interpreted as strict damping of every bare coordinate. It states only that the full homogeneous evolution cannot increase the vector norm. This identity is used later in the interval bounds for the phase family and the finite waveform neighborhood.

\suppnote{Same-waveform phase toggle and interval certificate}
\label{suppnote:phase_toggle}

The purpose of this note is to prove the causal pair used in Fig.~\ref{fig:phase_toggle}. We specify one power modulation and one detuning shape, obtain the two-EP control by a cyclic translation of that same detuning, prove that all scalar control resources match, certify the two windings independently of the cooling result, and then bound the final occupations with interval arithmetic.

\suppsubsection{Fixed power waveform and reference detuning}
\label{suppsec:phase_toggle_controls}

The power waveform is fixed first and is never translated. In physical units it is
\begin{equation}
P_L(t)
=
P_c+R_P\cos(\omega t),
\quad
P_c=4.7736340790737035,
\quad
R_P=1.1934085197684259,
\quad
\omega=\frac{2\pi}{T},
\quad
T=13.4.
\label{eq:supp_physical_power}
\end{equation}
The reference detuning is
\begin{equation}
\Delta_0(t)
=
\bar\Delta
+
\sum_{k=1}^{3}\left[a_k\cos(k\omega t)+b_k\sin(k\omega t)\right],
\quad
\bar\Delta=-2.5261340790737368,
\label{eq:supp_reference_detuning}
\end{equation}
with the effort-normalized coefficients
\begin{table}[t]
\centering
\begin{ruledtabular}
\begin{tabular}{crr}
$k$ & $a_k$ & $b_k$\\
\hline
1 & $-2.585803132791286$ & $0.185124160576254$\\
2 & $1.894600503929628$ & $-0.990552773303469$\\
3 & $0.691202628861658$ & $-0.717965845035285$
\end{tabular}
\end{ruledtabular}
\caption{Physical Fourier coefficients of the reference three-harmonic detuning waveform. Their cosine sum is zero to the certified precision, so the waveform starts and ends at $\bar\Delta$.\justifying}
\label{tab:supp_levelA_coefficients}
\end{table}
These coefficients define the reference waveform used throughout the causal comparison. They are fixed before the translated two-EP member is evaluated, and no separate waveform optimization is performed for the two topology classes.

Let
\begin{equation}
f(x)
=
\sum_{k=1}^{3}\left[a_k\cos(kx)+b_k\sin(kx)\right].
\label{eq:supp_f_definition}
\end{equation}
The non-enclosing control is $\Delta_{\controlclass_0}(t)=\bar\Delta+f(\omega t)$. A cyclic translation by an angular phase $\theta$ gives $\bar\Delta+f(\omega t+\theta)$. The endpoint condition for the translated waveform is $f(\theta)=0$. Interval root isolation of this trigonometric equation gives the nonzero phase
\begin{equation}
\theta_\ast=4.1915403811497277772,
\quad
t_\ast=8.9391985691121672081,
\label{eq:supp_theta_star}
\end{equation}
where $t_\ast=\theta_\ast/\omega$. This translated control is the selected two-EP trajectory $\controlclass_{12}$.

\suppsubsection{Exact resource matching}
\label{suppsec:phase_toggle_resources}

The key advantage of a cyclic translation is that the local waveform is unchanged. A full-period translation preserves the time average of the detuning fluctuation, its range, and every full-period integral that depends only on the translated waveform. In particular,
\begin{equation}
\int_0^T\dot\Delta_{\controlclass_0}^2(t)\,\dd t
=
\int_0^T\dot\Delta_{\controlclass_{12}}^2(t)\,\dd t
=50.
\label{eq:supp_effort_match}
\end{equation}
The maxima of $|\dot\Delta|$ and $|\ddot\Delta|$ are also translation invariants. Since $f(0)=f(\theta_\ast)=0$, both controls have the same endpoint $\bar\Delta$. The common certified geometry is
\begin{equation}
\Delta_{\min}
\in[-6.3153758246,-6.3153758245],
\quad
\Delta_{\max}
\in[1.2630862698,1.2630862699],
\label{eq:supp_common_range}
\end{equation}
with
\begin{equation}
\max|\dot\Delta|\le3.1777480085<4.6,
\quad
\max|\ddot\Delta|\le3.3939157833<4.2.
\label{eq:supp_common_derivative_bounds}
\end{equation}
Thus neither loop has access to a wider detuning range or faster modulation.

The topology is certified independently of the cooling result. The minimum distances from the $\controlclass_0$ loop to $({\rm EP}_1,{\rm EP}_2)$ are larger than $(0.16565,1.24951)$ in the physical control-plane metric used by the certificate, while the corresponding $\controlclass_{12}$ distances are larger than $(0.21569,0.40803)$. Both exceed the required clearance $\rho=0.08$. The winding numbers are
\begin{equation}
\controlclass_0:(w_1,w_2)=(0,0),
\quad
\controlclass_{12}:(w_1,w_2)=(1,1).
\label{eq:supp_levelA_windings}
\end{equation}
Because ${\rm EP}_1$ and ${\rm EP}_2$ exchange different sheet pairs, the second loop implements the composed three-branch cycle.

\suppsubsection{Validated endpoint dynamics}
\label{suppsec:phase_toggle_dynamics}

With the controls and topology fixed independently of the objective, only the cooling dynamics remain to be certified. The interval calculation uses the exact lossless linear dynamics and bounds the local Taylor residual, drift remainder, and accumulated propagation error on a complete time partition. The certified endpoint intervals are
\begin{equation}
\begin{aligned}
\controlclass_0:&\quad
n_m(T)\in[0.2128679579166367,\,0.2128686345528907],\\
\controlclass_{12}:&\quad
n_m(T)\in[0.0956793480200940,\,0.0956798218535326].
\end{aligned}
\label{eq:supp_levelA_certified_endpoints}
\end{equation}
Hence the guaranteed gap is at least
\begin{equation}
0.1171881360631042,
\label{eq:supp_levelA_gap}
\end{equation}
and the guaranteed reduction relative to the certified non-enclosing lower endpoint is at least $55.0520318840\%$. Direct full-covariance integration gives $0.212868296080886$ and $0.095679584896445$, respectively, inside the certified intervals.

This result is an explicit matched-pair theorem. It does not assert that the chosen $\controlclass_0$ control is globally optimal among all non-enclosing smooth waveforms.

\suppnote{Complete endpoint-anchored phase family}
\label{suppnote:phase_family}

The phase-toggle theorem compares two specific phases. To determine whether those phases are isolated, this note constructs a one-parameter family that scans the full phase circle while keeping the endpoint, mean, and effort fixed. We then certify every admissible phase sector and bound the dynamics over the complete conservative hulls.

\suppsubsection{Family construction}
\label{suppsec:anchored_family}

A pure translation keeps all full-period shape resources fixed, but for a general phase it does not return to the prescribed endpoint. To scan the complete phase circle without giving up the endpoint, mean, or effort constraints, we therefore add the smallest anchoring correction needed for those equalities. With $x=\omega t$ and $f$ from Eq.~\eqref{eq:supp_f_definition}, define
\begin{equation}
F_\theta(x)
=
f(x+\theta)-f(\theta)\cos x.
\label{eq:supp_Ftheta}
\end{equation}
Since $F_\theta(0)=F_\theta(2\pi)=0$ and both terms have zero full-period mean, the waveform
\begin{equation}
\Delta_\theta(t)
=
\bar\Delta+s_\theta F_\theta(\omega t)
\label{eq:supp_anchored_control}
\end{equation}
has the common endpoint and mean for every $\theta$. The positive scale factor is chosen as
\begin{equation}
s_\theta
=
\sqrt{\frac{50}{\int_0^T\left[\frac{\dd}{\dd t}F_\theta(\omega t)\right]^2\dd t}},
\label{eq:supp_anchored_scale}
\end{equation}
so the effort is exactly $50$. At $\theta=0$ and $\theta=\theta_\ast$, $f(\theta)=0$ and $s_\theta=1$, so Eq.~\eqref{eq:supp_anchored_control} reproduces the two phase-toggle controls exactly.

\suppsubsection{Exhaustive phase cover and topology sectors}
\label{suppsec:phase_cover}

We next determine where this family is physically admissible. The full interval $0\le\theta<2\pi$ is covered with interval arithmetic. The complement of the candidate sectors is adaptively divided into $911$ interval boxes. Every such box is rejected because at least one required detuning-extremum band is violated. This leaves four conservative hulls,
\begin{table}[t]
\centering
\begin{ruledtabular}
\begin{tabular}{lcc}
sector & phase interval & winding $(w_1,w_2)$\\
\hline
$\controlclass_0^{(a)}$ & $[-0.00082,0.00083]$ modulo $2\pi$ & $(0,0)$\\
$\controlclass_{12}^{-}$ & $[1.72036,1.72142]$ & $(-1,-1)$\\
$\controlclass_{12}^{+}$ & $[4.19108,4.19199]$ & $(1,1)$\\
$\controlclass_0^{(b)}$ & $[5.70434,5.70638]$ & $(0,0)$
\end{tabular}
\end{ruledtabular}
\caption{Conservative phase hulls, with phase intervals in radians, containing every fully admissible member of the endpoint-anchored family. No fully admissible single-EP sector remains outside these four hulls.\justifying}
\label{tab:supp_phase_hulls}
\end{table}
Each complete hull is then checked for the slew and acceleration bounds, exceptional-point clearance, and fixed winding. The smallest certified clearance over all four hulls remains above the required $\rho=0.08$.

\begin{figure}[t]
\centering
\includegraphics[width=0.72\linewidth]{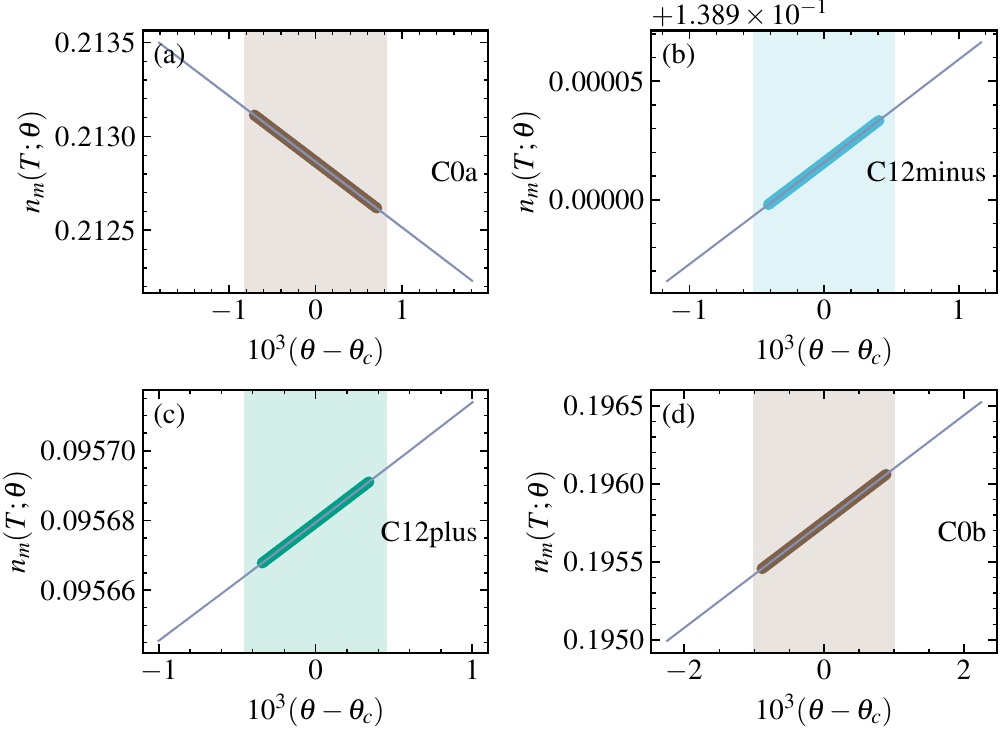}
\caption{Zooms of the four conservative phase hulls retained by the exhaustive interval cover. The dense curves are numerical guides; the hull boundaries and admissibility statements come from interval bounds.\justifying}
\label{fig:supp_phase_hull_zooms}
\end{figure}

The four panels resolve the sectors listed in Table~\ref{tab:supp_phase_hulls}: Fig.~\figpanel{fig:supp_phase_hull_zooms}{a} is the first non-enclosing sector, Fig.~\figpanel{fig:supp_phase_hull_zooms}{b} is the negative-orientation two-EP sector, Fig.~\figpanel{fig:supp_phase_hull_zooms}{c} is the positive-orientation two-EP sector containing the main phase-toggle point, and Fig.~\figpanel{fig:supp_phase_hull_zooms}{d} is the second non-enclosing sector. The smooth curves are dense numerical evaluations included only to show the local shape. The displayed phase hulls and all admissibility statements come from interval bounds, so the proof does not depend on the sampling density.

\suppsubsection{Family-wide dynamical ordering}
\label{suppsec:phase_family_dynamics}

We now propagate the dynamics over these sectors. To avoid using a hidden numerical feasibility selection, the dynamics are bounded over the complete conservative hulls, including points inside those hulls that may themselves fail one of the extrema conditions. This only weakens the bounds and therefore makes the final ordering conservative. The resulting enclosures are
\begin{table}[t]
\centering
\begin{ruledtabular}
\begin{tabular}{lc}
sector & certified enclosure of $n_m(T)$\\
\hline
$\controlclass_0^{(a)}$ & $[0.1708195629,0.2615160521]$\\
$\controlclass_{12}^{-}$ & $[0.1281776831,0.1512866265]$\\
$\controlclass_{12}^{+}$ & $[0.0778595758,0.1188264069]$\\
$\controlclass_0^{(b)}$ & $[0.1589328792,0.2393347125]$
\end{tabular}
\end{ruledtabular}
\caption{Conservative occupation bounds over the complete phase hulls. Because the bounds hold on the full hulls, they also hold on the admissible subsets contained within them.\justifying}
\label{tab:supp_phase_family_objectives}
\end{table}
Combining the two non-enclosing hulls and the two two-EP hulls gives Eq.~\eqref{eq:main_levelB_all_to_all}. The guaranteed all-to-all gap is at least $0.0076462527$. Restricting to the selected positive orientation gives the stronger bound
\begin{equation}
\sup_{\theta\in\controlclass_{12}^{+}}n_m(T;\theta)
\le0.1188264069
<0.1589328792
\le
\inf_{\theta\in\controlclass_0}n_m(T;\theta).
\label{eq:supp_plus_orientation_ordering}
\end{equation}
The exhaustive family theorem is therefore stronger than a local continuity statement around the phase-toggle pair, but it remains a theorem for the anchored one-parameter family rather than for all smooth controls.

\suppnote{Robustness to waveform shape and smooth perturbations}
\label{suppnote:waveform_robustness}

The complete phase family still inherits one basic Fourier shape. This note asks a different question: does the ordering survive when the waveform itself changes? We first test several predeclared redistributions of the spectral effort, then repeat the comparison with a periodic cubic spline, and finally prove a finite neighborhood of arbitrary smooth perturbations within the set of exactly matched controls.

\suppsubsection{Predeclared Fourier shape changes}
\label{suppsec:shape_robustness}

To avoid choosing waveform shapes after looking at the cooling result, the alternative Fourier bank is fixed by spectral effort fractions before $n_m(T)$ is evaluated. For
\begin{equation}
f(x)
=
\sum_{k=1}^{K}r_k\cos(kx-\varphi_k),
\label{eq:supp_shape_amplitude_form}
\end{equation}
the fraction $p_k$ of the derivative-effort budget in harmonic $k$ is fixed through
\begin{equation}
\frac{T\omega^2}{2}k^2r_k^2=50p_k,
\quad
\sum_kp_k=1.
\label{eq:supp_spectral_effort_fraction}
\end{equation}
The phases and the two anchored comparison points are selected using only extrema matching, topology, exceptional-point clearance, and slew/acceleration slack. The cooling objective is not used for this selection.

The reference effort fractions are $(0.19800043,0.53863951,0.26336006)$. We test three changes:
\begin{table}[t]
\centering
\begin{ruledtabular}
\begin{tabular}{lccc}
shape & $n_m^{\controlclass_0}(T)$ & $n_m^{\controlclass_{12}}(T)$ & guaranteed reduction\\
\hline
K3 low shift & $[0.1643022101,0.1643028719]$ & $[0.1099559412,0.1099562107]$ & $33.076852\%$\\
K3 high shift & $[0.2430669950,0.2430671227]$ & $[0.0991875370,0.0991879836]$ & $59.193150\%$\\
K4, 5\% new harmonic & $[0.2380908974,0.2380910092]$ & $[0.0894971493,0.0894976607]$ & $62.410297\%$
\end{tabular}
\end{ruledtabular}
\caption{Interval-certified waveform-shape robustness. The low-shift shape transfers ten percentage points of effort from harmonic 2 to 1, the high-shift shape transfers ten points from harmonic 2 to 3, and the K4 shape assigns five percent of the effort to a new fourth harmonic.\justifying}
\label{tab:supp_shape_robustness}
\end{table}
All accepted pairs have windings $(0,0)$ and $(1,1)$ and satisfy the same endpoint, mean, duration, fixed power waveform, and effort constraints.

\begin{figure}[t]
\centering
\includegraphics[width=0.72\linewidth]{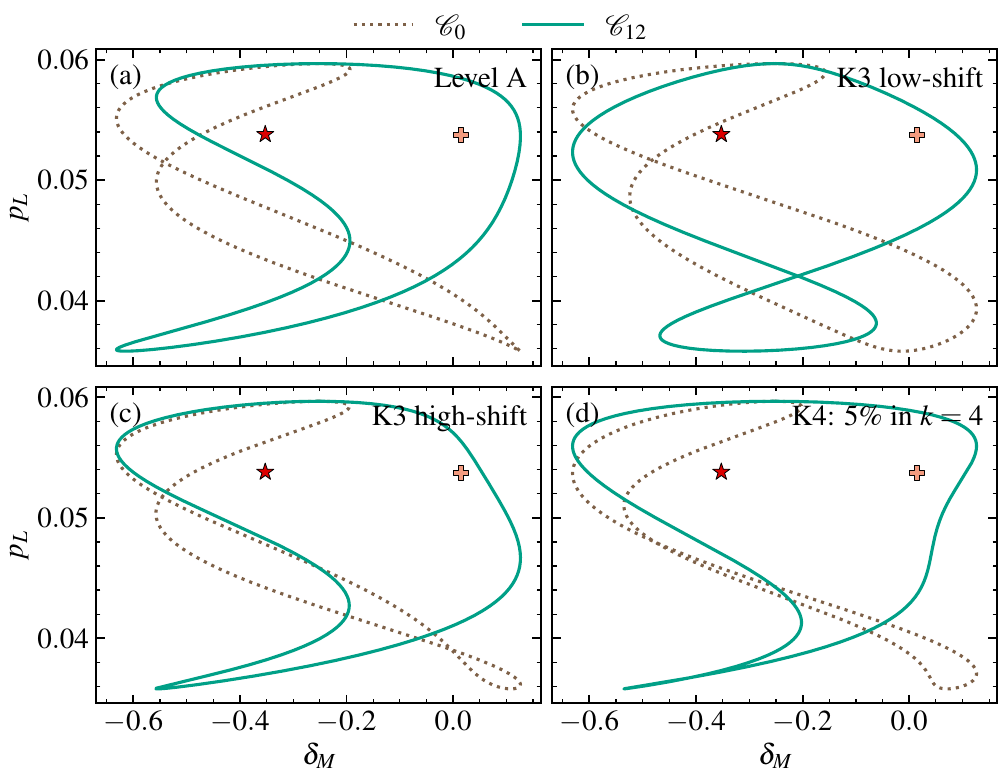}
\caption{Control-plane paths for the phase-toggle reference and the three interval-certified waveform-shape tests. Each pair contains a non-enclosing and a two-EP control selected without using the cooling objective.\justifying}
\label{fig:supp_shape_control_paths}
\end{figure}

Figure~\figpanel{fig:supp_shape_control_paths}{a} shows the phase-toggle reference pair. Figures~\figpanel{fig:supp_shape_control_paths}{b} and \figpanel{fig:supp_shape_control_paths}{c} show the two opposite redistributions within the three-harmonic bank, and Fig.~\figpanel{fig:supp_shape_control_paths}{d} shows the four-harmonic test. In every panel the two paths satisfy the same local resources but realize $(0,0)$ and $(1,1)$ windings. The persistence of the topology change across visibly different loops is part of the geometry check; the cooling values are certified separately.

\begin{figure}[t]
\centering
\includegraphics[width=0.72\linewidth]{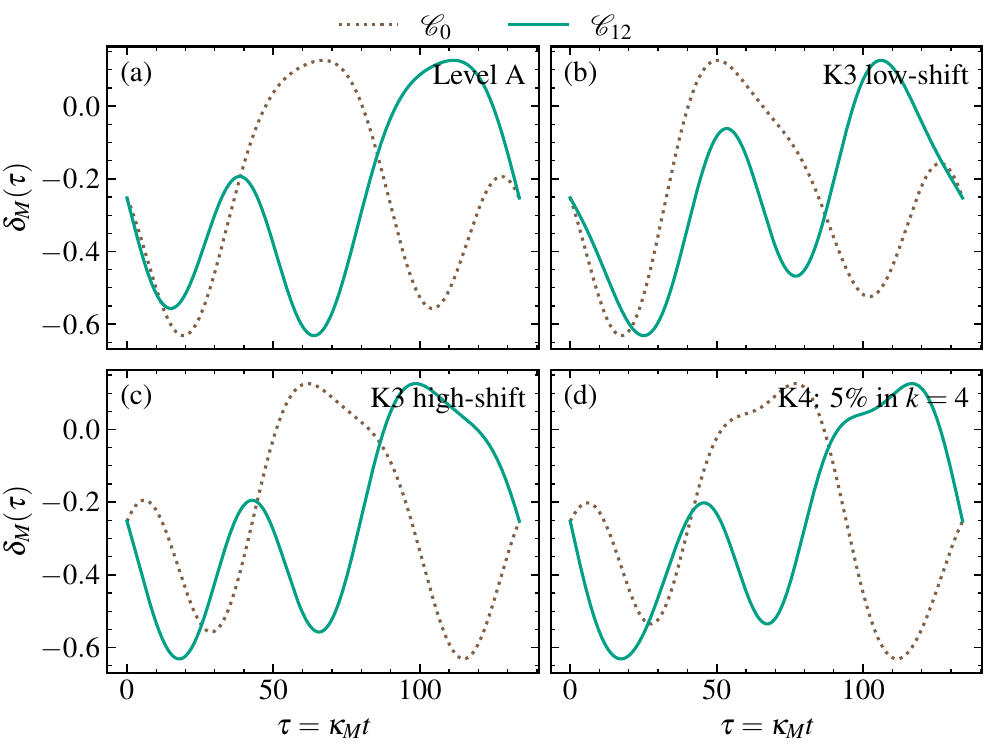}
\caption{Detuning waveforms for the phase-toggle reference and the three discrete shape-robustness tests. The shapes redistribute the same derivative-effort budget rather than increasing it.\justifying}
\label{fig:supp_shape_detunings}
\end{figure}

The corresponding time-domain shapes are shown in Fig.~\ref{fig:supp_shape_detunings}. Figure~\figpanel{fig:supp_shape_detunings}{a} is the reference waveform. Figures~\figpanel{fig:supp_shape_detunings}{b} and \figpanel{fig:supp_shape_detunings}{c} visibly move spectral weight toward lower and higher harmonics, while Fig.~\figpanel{fig:supp_shape_detunings}{d} introduces the new fourth harmonic. The purpose of the figure is to make clear that robustness is being tested against genuine shape changes, not against small numerical variations of one curve.

\suppsubsection{Independent periodic-spline cross-check}
\label{suppsec:spline_crosscheck}

The previous tests remain Fourier based, so we next change the mathematical representation itself. As a non-Fourier check, the reference detuning is sampled on $20$ equally spaced knots and replaced by a periodic cubic spline. The mean and effort are re-normalized without using the cooling result, and the phase choices are again made from geometry alone. The resulting numerical endpoints are
\begin{equation}
 n_m^{\controlclass_0}(T)=0.212765316007,
\quad
 n_m^{\controlclass_{12}}(T)=0.095663705425,
\label{eq:supp_spline_endpoints}
\end{equation}
corresponding to a $55.0379\%$ reduction. Independent full-covariance and reduced-response integrations agree at the $10^{-12}$ level. This spline result is a high-precision numerical representation check; it is not included in the interval theorem.

\begin{figure}[t]
\centering
\includegraphics[width=0.70\linewidth]{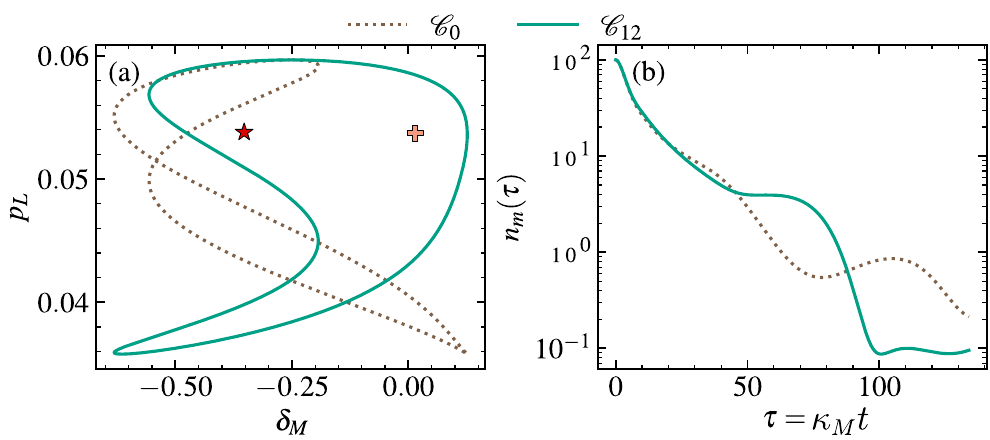}
\caption{Independent periodic-cubic-spline cross-check with $20$ knots. The spline representation reproduces the same non-enclosing versus two-EP hierarchy without using the Fourier basis.\justifying}
\label{fig:supp_spline_crosscheck}
\end{figure}

Figure~\figpanel{fig:supp_spline_crosscheck}{a} shows that the spline pair retains the non-enclosing and two-EP control-plane topology. Figure~\figpanel{fig:supp_spline_crosscheck}{b} shows the corresponding cooling traces and their final separation. Agreement between the full covariance and reduced-response integrations at the $10^{-12}$ level makes this a strong independent numerical cross-check, but we keep it logically separate from the interval-certified statements above.

\suppsubsection{Finite neighborhood with exactly matched resources}
\label{suppsec:open_neighborhood}

A finite list of alternative waveforms still leaves open the possibility that the ordering disappears under arbitrarily small generic deformations. We therefore finish the robustness analysis with a continuous neighborhood. Writing the detuning perturbation away from either phase-toggle reference as $\delta\Delta$, we consider arbitrary smooth perturbations that preserve the endpoint, mean, and effort exactly and satisfy
\begin{equation}
\|\delta\Delta\|_\infty\le1.5\times10^{-3},
\quad
\|\delta\dot\Delta\|_\infty\le0.20,
\quad
\|\delta\ddot\Delta\|_\infty\le0.20.
\label{eq:supp_neighborhood_radii}
\end{equation}
Interval bounds show that every control in these neighborhoods remains inside the detuning-extrema bands and below the slew and acceleration limits. A straight continuous interpolation from any perturbed waveform back to its phase-toggle reference stays clear of both exceptional points, so the winding cannot change.

The dynamical bound uses Eq.~\eqref{eq:supp_nonexpansive_identity}. For the exact two-time propagator $X(T,s)$, the endpoint mechanical occupation can be written as
\begin{equation}
n_m(T)
=
n_m(0)|X_{mm}(T,0)|^2
+\gamma_m n_m^{\rm th}\int_0^T|X_{mm}(T,s)|^2\,\dd s.
\label{eq:supp_reduced_endpoint_formula}
\end{equation}
Comparing the perturbed and reference propagators with the variation-of-constants (Duhamel) formula and using non-expansiveness gives a uniform state perturbation below $1.7364\times10^{-3}$ at the full radius in Eq.~\eqref{eq:supp_neighborhood_radii}. Combining this with the validated phase-toggle state bounds gives the following result. Denoting arbitrary members of the two neighborhoods by $\widetilde{\controlclass}_0$ and $\widetilde{\controlclass}_{12}$,
\begin{equation}
 n_m^{\widetilde{\controlclass}_0}(T)\ge0.1993737215,
\quad
 n_m^{\widetilde{\controlclass}_{12}}(T)\le0.1029231965.
\label{eq:supp_open_neighborhood_result}
\end{equation}
The guaranteed gap is therefore at least $0.0964505250$ throughout the complete stated neighborhoods. Figure~\figpanel{fig:waveform_robustness}{b} of the main text plots this certified gap as the detuning radius is increased from zero to its full value. The neighborhood is restricted to controls that keep the prescribed resources exactly matched; it is not a statement about unrestricted perturbations that change those resources.

\suppnote{Fixed-control validation beyond the rotating-wave approximation}
\label{suppnote:bdg}

The topology certificates above are formulated for the time-local RWA drift. This note tests the omitted counter-rotating heating processes without giving any control a second chance to optimize. The RWA controls are held fixed, the dynamical model is enlarged, and the resulting endpoint changes can therefore be attributed only to the additional physical terms.

The normalized linearized Hamiltonian separates naturally into the RWA part used for the static exceptional-point geometry and a counter-rotating part,
\begin{equation}
\frac{H_{\rm lin}(\tau)}{\hbar\kappa_M}
=
\frac{H_{\rm RWA}(\tau)}{\hbar\kappa_M}
+
\frac{H_{\rm CR}(\tau)}{\hbar\kappa_M},
\label{eq:supp_bdg_hamiltonian_split}
\end{equation}
where
\begin{equation}
\frac{H_{\rm CR}(\tau)}{\hbar\kappa_M}
=
g(\tau)e^{2i\Omega_m\tau}a^\dagger m^\dagger
+g^\ast(\tau)e^{-2i\Omega_m\tau}am,
\quad
\Omega_m=10.
\label{eq:supp_counterrotating_hamiltonian}
\end{equation}
These terms create or annihilate a photon--phonon pair and therefore describe the Stokes processes omitted by the RWA\@. Because their oscillating phase makes the enlarged drift explicitly time dependent even at fixed control coordinates, we use them as a dynamical validation and do not redefine the static EP topology in this explicitly time-dependent model.

We use the standard doubled Bogoliubov--de Gennes (BdG) representation. For the vector
\begin{equation}
\bm w=(a,m,c,a^\dagger,m^\dagger,c^\dagger)^T,
\label{eq:supp_bdg_vector}
\end{equation}
the homogeneous equation is
\begin{equation}
\frac{\dd\bm w}{\dd\tau}
=
M_{\rm BdG}(\tau)\bm w,
\quad
M_{\rm BdG}(\tau)
=
\begin{pmatrix}
A_{\rm full}(\tau)&B(\tau)\\
B^\ast(\tau)&A_{\rm full}^\ast(\tau)
\end{pmatrix},
\label{eq:supp_bdg_drift}
\end{equation}
with
\begin{equation}
B(\tau)
=
-ig(\tau)e^{2i\Omega_m\tau}
\begin{pmatrix}
0&1&0\\
1&0&0\\
0&0&0
\end{pmatrix}.
\label{eq:supp_bdg_block}
\end{equation}
The doubled covariance $C_{ij}=\langle w_iw_j\rangle$ obeys
\begin{equation}
\frac{\dd C}{\dd\tau}
=
M_{\rm BdG}C+CM_{\rm BdG}^{T}+D_{\rm BdG}.
\label{eq:supp_bdg_covariance}
\end{equation}
Because the auxiliary cavity is lossless, its rows and columns in the diffusion blocks carry no reservoir term. With the optical baths in vacuum and the mechanical bath at occupation $100$,
\begin{equation}
D_+
=
{\rm diag}\!\left(1,\widetilde\gamma_m(n_m^{\rm th}+1),0\right),
\quad
D_-
=
{\rm diag}\!\left(0,\widetilde\gamma_m n_m^{\rm th},0\right),
\label{eq:supp_bdg_diffusion_blocks}
\end{equation}
and $D_{\rm BdG}=\left(\begin{smallmatrix}0&D_+\\D_-&0\end{smallmatrix}\right)$.

Setting $B(\tau)=0$ must reproduce the normally ordered RWA covariance calculation. Across all fixed controls listed below, the maximum endpoint discrepancy is $2.41\times10^{-12}$. Restoring $B(\tau)$ gives
\begin{table}[t]
\centering
\begin{ruledtabular}
\begin{tabular}{lccc}
fixed pair & control & RWA & full BdG\\
\hline
phase toggle & $\controlclass_0$ & $0.2128682961$ & $0.2455144837$\\
phase toggle & $\controlclass_{12}$ & $0.0956795849$ & $0.0973701718$\\
phase-family representative & $\controlclass_0^{(b)}$ & $0.1957584525$ & $0.2228587611$\\
phase-family representative & $\controlclass_{12}^{-}$ & $0.1389158137$ & $0.1457111707$\\
K4 shape test & $\controlclass_0$ & $0.2380909533$ & $0.2717761419$\\
K4 shape test & $\controlclass_{12}$ & $0.0894974050$ & $0.0964054328$
\end{tabular}
\end{ruledtabular}
\caption{Fixed-control beyond-RWA validation. No waveform is reoptimized after the counter-rotating block is restored. The full-Bogoliubov reductions within the three pairs are $60.3404\%$, $34.6173\%$, and $64.5276\%$, respectively.\justifying}
\label{tab:supp_bdg_endpoints}
\end{table}

\begin{figure}[t]
\centering
\includegraphics[width=0.72\linewidth]{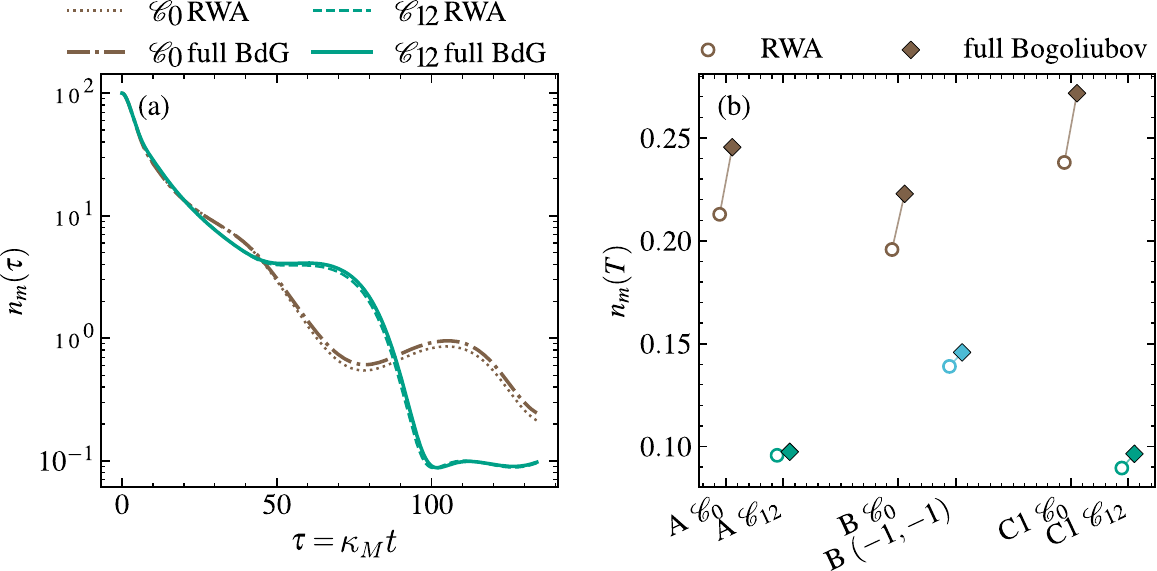}
\caption{Fixed-control Bogoliubov validation. Panel (a) compares the RWA and full-Bogoliubov time traces for the phase-toggle pair. Panel (b) compares RWA and full-Bogoliubov endpoints for the phase-toggle pair and representative phase-family and waveform-shape controls.\justifying}
\label{fig:supp_bdg_validation}
\end{figure}

Figure~\figpanel{fig:supp_bdg_validation}{a} compares the full time traces of the phase-toggle pair before and after the counter-rotating block is restored. The correction raises the non-enclosing endpoint much more strongly than the two-EP endpoint, so the ordering not only survives but remains large. Figure~\figpanel{fig:supp_bdg_validation}{b} repeats the endpoint comparison for the phase-toggle pair, a representative pair from the complete anchored phase family, and the four-harmonic shape test. All three fixed pairs preserve the two-EP advantage. No point in either panel results from a Bogoliubov-level reoptimization.

The phase-toggle source decomposition is also stable. The exact RWA and full-Bogoliubov endpoint splits are
\begin{table}[t]
\centering
\begin{ruledtabular}
\begin{tabular}{lcccc}
 & \multicolumn{2}{c}{RWA} & \multicolumn{2}{c}{Full BdG}\\
control & initial & bath & initial & bath\\
\hline
$\controlclass_0$ & $0.1329698731$ & $0.0798984230$ & $0.1638392930$ & $0.0816751907$\\
$\controlclass_{12}$ & $0.0288493774$ & $0.0668302075$ & $0.0288592927$ & $0.0685108791$
\end{tabular}
\end{ruledtabular}
\caption{Endpoint mechanical occupation split into the contribution propagated from the initial covariance and the contribution generated by reservoir noise. In the RWA, $88.85\%$ of the $\controlclass_0-\controlclass_{12}$ gap comes from the initial component; in the full Bogoliubov calculation the fraction is $91.11\%$.\justifying}
\label{tab:supp_bdg_source_split}
\end{table}

\begin{figure}[t]
\centering
\includegraphics[width=0.70\linewidth]{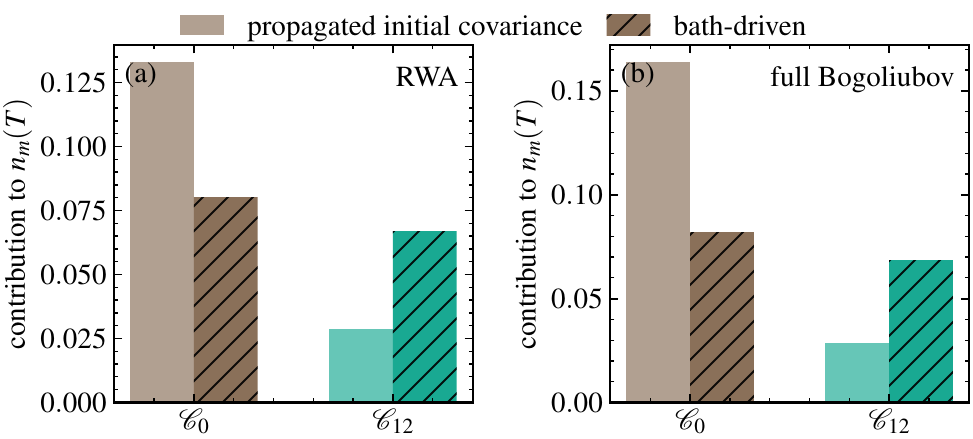}
\caption{Source-resolved phase-toggle endpoints in the RWA and full Bogoliubov dynamics. The dominant difference between the non-enclosing and two-EP controls comes from the amount of the initial mechanical excitation that survives to the endpoint.\justifying}
\label{fig:supp_bdg_source_split}
\end{figure}

Figure~\figpanel{fig:supp_bdg_source_split}{a} shows the RWA source split and Fig.~\figpanel{fig:supp_bdg_source_split}{b} shows the full-Bogoliubov split. In both models the main separation comes from the excitation propagated from the initial mechanical state, while the bath-generated contribution changes more modestly. The counter-rotating terms therefore do not replace the mechanism identified in the RWA analysis.

The finite-neighborhood result of Note~\ref{suppnote:waveform_robustness} is not extended here to a full-Bogoliubov interval theorem. The present calculation validates the phase-toggle center pair and representative members of the broader tests with the controls held fixed.

\suppnote{Lossless coherent-routing mechanism}
\label{suppnote:mechanism}

The previous notes establish the causal comparison, its robustness, and its survival beyond the RWA\@. We now use the lossless auxiliary limit to identify the physical route. Because the auxiliary cavity has no dissipative channel, any occupation that enters it must later return coherently or remain stored there. This lets us distinguish coherent spectral buffering from irreversible removal by the main cavity.

\suppsubsection{Source decomposition and excitation balance}
\label{suppsec:mechanism_balance}

Let $X(\tau,\sigma)$ be the RWA propagator generated by Eq.~\eqref{eq:main_full_drift}. Linearity gives the exact covariance solution
\begin{equation}
N(\mathcal T)
=
X^\ast(\mathcal T,0)N(0)X^T(\mathcal T,0)
+
\int_0^{\mathcal T}
X^\ast(\mathcal T,\sigma)D_{\rm th}X^T(\mathcal T,\sigma)\,\dd\sigma.
\label{eq:supp_covariance_source_solution}
\end{equation}
This defines an exact split into propagated initial excitation and reservoir-generated excitation. The numerical endpoint values for the phase-toggle pair are given in Table~\ref{tab:supp_bdg_source_split}.

For real $g(\tau)$ and $j$, define the dimensionless coherent currents
\begin{equation}
\controlclassJ_{m\to a}(\tau)
=
2g(\tau)\,\operatorname{Im}\langle a^\dagger m\rangle,
\quad
\controlclassJ_{c\to a}(\tau)
=
2j\,\operatorname{Im}\langle a^\dagger c\rangle.
\label{eq:supp_coherent_currents}
\end{equation}
Positive current denotes transfer into the main cavity. The bare occupations satisfy
\begin{equation}
\begin{aligned}
\dot n_a&=-n_a+\controlclassJ_{m\to a}+\controlclassJ_{c\to a},\\
\dot n_m&=\widetilde\gamma_m(n_m^{\rm th}-n_m)-\controlclassJ_{m\to a},\\
\dot n_c&=-\controlclassJ_{c\to a},
\end{aligned}
\label{eq:supp_occupation_balance}
\end{equation}
where the dot now denotes differentiation with respect to $\tau$. Adding the three equations gives
\begin{equation}
\frac{\dd}{\dd\tau}(n_a+n_m+n_c)
=
\widetilde\gamma_m(n_m^{\rm th}-n_m)-n_a.
\label{eq:supp_total_balance}
\end{equation}
The coherent currents cancel from the total balance. More importantly, there is no auxiliary dissipative term. The auxiliary cavity redistributes excitation but cannot remove it.

To quantify repeated coherent exchange, we use
\begin{equation}
Q_{a\leftrightarrow c}^{\rm gross}(\tau)
=
\int_0^\tau|\controlclassJ_{c\to a}(s)|\,\dd s,
\quad
L_M(\tau)
=
\int_0^\tau n_a(s)\,\dd s.
\label{eq:supp_gross_exchange_loss}
\end{equation}
For the phase-toggle pair, $\max n_c$ is $10.77809$ for $\controlclass_0$ and $11.93194$ for $\controlclass_{12}$; the completed gross optical exchange is $21.10645$ and $23.63105$, while the main-cavity loss is $100.53495$ and $100.88978$. The auxiliary loss is exactly zero for both controls. These quantities are shown in Fig.~\ref{fig:mechanism} of the main text. Figure~\figpanel{fig:mechanism}{a} demonstrates substantial temporary auxiliary occupation, Fig.~\figpanel{fig:mechanism}{b} shows the current reversing direction, Fig.~\figpanel{fig:mechanism}{c} accumulates the repeated bidirectional exchange, and Fig.~\figpanel{fig:mechanism}{d} shows that the irreversible optical loss occurs only through the main cavity. The similar final main-cavity losses, despite very different final mechanical occupations, show that timing rather than total extracted amount is the decisive difference.

\suppsubsection{Tracked spectral composition}
\label{suppsec:branch_composition}

The occupation balance tells us where excitation can flow, but it does not by itself show how the spectrum reorganizes that flow. The coherent-buffer picture can also be read from the instantaneous eigenvectors. Let $r_j(\tau)$ be a continuously tracked right eigenvector of the drift. We normalize each right eigenvector in the Euclidean norm and define its bare-mode weights by
\begin{equation}
W_{\mu,j}(\tau)
=
\frac{|r_{\mu,j}(\tau)|^2}
{|r_{a,j}(\tau)|^2+|r_{m,j}(\tau)|^2+|r_{c,j}(\tau)|^2},
\quad
\mu\in\{a,m,c\}.
\label{eq:supp_branch_weights}
\end{equation}
To continue the branch labels, we compute the left and right eigenvectors at neighboring time steps and choose the assignment with the largest overlap, rather than re-sorting the eigenvalues independently at each step. Around the two-EP loop, the three branch identities undergo the composed cycle implied by the two exceptional points, and their bare-mode compositions are redistributed accordingly.

\begin{figure}[t]
\centering
\includegraphics[width=0.62\linewidth]{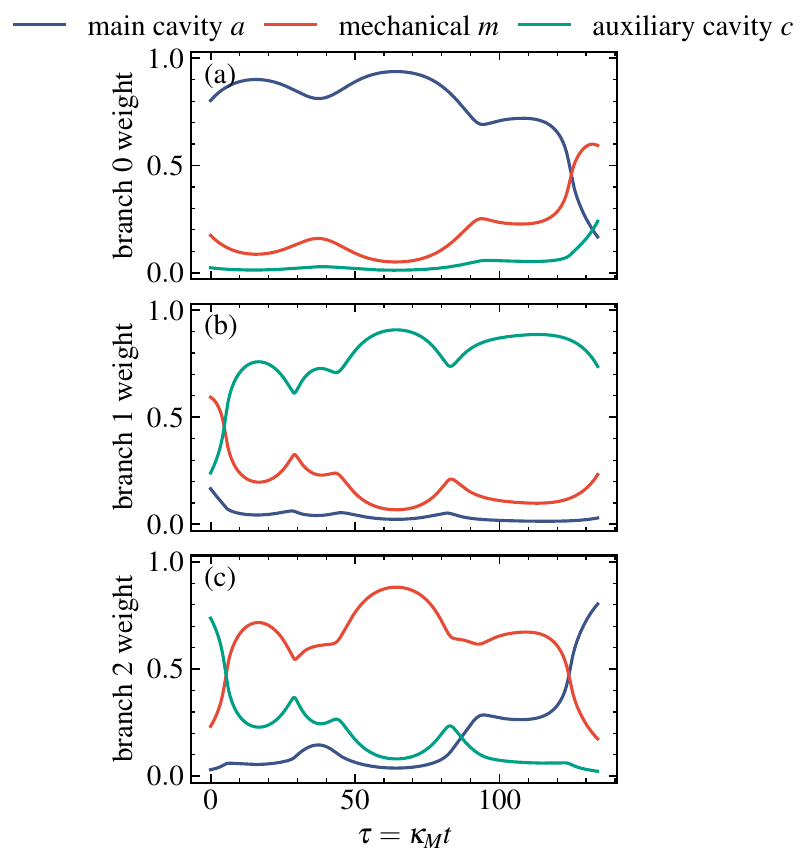}
\caption{Bare-mode composition of the three continuously tracked drift branches along the phase-toggle two-EP control. Each panel shows the normalized main-cavity, mechanical, and auxiliary weights of one branch. The plot diagnoses the available spectral routing structure; it is not an assumption of adiabatic state following.\justifying}
\label{fig:supp_branch_composition}
\end{figure}

Figures~\figpanel{fig:supp_branch_composition}{a}, \figpanel{fig:supp_branch_composition}{b}, and \figpanel{fig:supp_branch_composition}{c} follow branches $0$, $1$, and $2$, respectively, around the same two-EP loop. In each panel the three curves give the main-cavity, mechanical, and auxiliary components of that one continuously tracked eigenvector. The dominant bare-mode character is exchanged as the loop proceeds, making visible the spectral route that is absent in a non-enclosing cycle. These weights describe instantaneous eigenvector composition, not state populations; no adiabatic-following assumption is used in the dynamical calculation.

The channel and branch diagnostics support one consistent interpretation. The lossless auxiliary mode provides storage and a third hybrid branch, the two exceptional points connect different sheet pairs, and the two-EP cycle changes the order in which long-lived and main-cavity-like character appear along the connected route. The main cavity then supplies the irreversible loss. This mechanism explains why similar total loss can lead to different final mechanical occupations: the timing of transfer and loss matters, not only their cycle-integrated amounts.

\end{document}